# Direct Meissner Effect Observation of Superconductivity in Compressed $H_2S$

*Xiaoli Huang[1†], Xin Wang[1†], Defang Duan[1], Bertil Sundqvist[1,2], Xin Li[1], Yanping Huang[1], Fangfei Li[1], Qiang Zhou[1], Bingbing Liu[1], and Tian Cui[1]\**

[1]State Key Laboratory of Superhard Materials, College of Physics, Jilin University, Changchun 130012, People's Republic of China.

[2]Department of Physics, Umeå University, SE-90187 Umeå, Sweden

\*Corresponding author, E-mail: cuitian@jlu.edu.cn, Tel. / Fax: +86-431-85168825

[†]These authors contribute equally to this work

**Abstract**

Recently, an extremely high superconducting temperature ($T_c$) of ~200 K has been reported in the sulfur hydride system above 100 GPa. This result is supported by theoretical predictions and verified experimentally. The crystal structure of the superconducting phase was also identified experimentally, confirming the theoretically predicted structure as well as a decomposition mechanism from $H_2S$ to $H_3S$+S. Even though nuclear resonant scattering has been successfully used to provide magnetic evidence for a superconducting state, a direct measurement of the important Meissner effect is still lacking. Here we report *in situ* alternating-current magnetic susceptibility measurements on compressed $H_2S$ under high pressures. It is shown that superconductivity suddenly appears at 117 GPa and that $T_c$ reaches 183 K at 149 GPa before decreasing monotonically with a further increase in pressure. This evolution agrees with both theoretical calculations and earlier experimental measurements. The idea of conventional high temperature superconductivity in hydrogen-dominant compounds has thus been realized in the sulfur hydride system under hydrostatic pressure, opening further exciting perspectives for possibly realizing room temperature superconductivity in hydrogen-based compounds.

Since the discovery of superconductivity slightly more than a century ago (*1*), the quest for ever higher critical onset temperatures, $T_c$, has remained a great challenge and an important topic for experimental and theoretical research. Ashcroft proposed that hydrogen-dominated materials with high Debye temperatures might show high-$T_c$ superconductivity under pressures lower than those expected for metallic hydrogen (*2, 3*). This proposal has been supported by numerous calculations, for example, ref. *4 – 6*. Some hydrides have been predicted theoretically as good candidates to realize high temperature superconductivity, but in spite of great experimental efforts the field showed no significant breakthroughs for a long time.

Very recently, however, the highest known values for the superconducting transition temperature $T_c$ were observed in a newly discovered hydrogen sulfide system with a record $T_c$ of 200 K (*6*), exceeding the highest values found for the cuprates which have held the record for nearly 30 years. The high $T_c$ was indicated by a sharp drop of the resistance to zero and confirmed by a strong isotope effect in $D_2S$ (*7, 8*). Similar experiments performed by Shimizu *et al.* further verified a resistance drop in the sulfur hydride system above 90 GPa (*9*), the predicted structures and a predicted decomposition of $H_2S$ to $H_3S+S$ (*10*). These results have kindled an intense search for room-temperature superconductivity in hydrogen-dominated materials.

The Meissner effect is considered the most important feature to prove the presence of a superconducting state, although Troyan *et al.* reported the observation of an expulsion of the magnetic field in compressed $H_2S$ at 153 GPa (*11*). Here we have used a highly sensitive magnetic susceptibility technique adapted for a megabar-pressure diamond anvil cell to explore and demonstrate the existence of a superconducting Meissner effect in sulfur hydrides and to verify the superconductivity phase diagram of the material.

The details of the loading procedure for the $H_2S$ sample are described in Methods. With different thermodynamic routes, the sample undergoes different phase transitions, especially at different temperatures. The sample was compressed to at least 100 GPa at $T < 100$ K after loading. The magnetic susceptibility measurement setup in the diamond anvil cell is shown in

Fig. 1A. The alternating-current susceptibility was detected with a lock-in amplifier (LIA). The target sample was loaded into the hole of the non-magnetic gasket at about 2.0 GPa and 200 K, and the sample at 155 GPa and 300 K is shown in Fig. 1B. The signal coil is wound around one of the diamond anvils with a compensating coil connected in opposition (Fig. 1C). The basic principles of the magnetic-susceptibility technique for measuring the superconductivity are described in previous works (*12*, *13*). We have extended this technique to study samples in the megabar pressure range, using a technique similar with that used by Deemyad *et al.* (*13*). The value of $T_c$ is taken to be the onset of superconductivity, defined by the intersection of a line drawn through the steep slope of the curve and the region of zero slope above the transition.

Two extensive experimental runs have been performed and are summarized in Fig. 2. In the first experiment the sample was compressed up to 149 GPa, above which the diamond failed. Below 100 GPa, no superconducting transition was observed in the sample, but at 117 GPa a characteristic transition into the superconducting state was observed as shown in Fig. 2A. The $T_c$, identified as the onset of the steep drop, is found to be 38 K. At 130 GPa, the superconducting onset temperature is 55 K (Fig. 2B). The pressure was then increased to 149 GPa at 300 K, where the sample was kept for about seven days, waiting out the possible transition kinetics in $H_3S$ as reported in previous works (*7*, *9*). When the temperature was finally decreased at this pressure the superconducting transition was observed at 183 K (Fig. 2C), which is qualitatively compatible with our theoretical predictions and, within the experimental uncertainties, with previous electrical resistance measurements (*6*, *7*). In the second run, the pressure was increased monotonically from 155 GPa to 171 GPa. At 155 GPa, a well-defined and reproducible superconducting signal was triggered at about 168 K (Fig. 2D). However, when the pressure was increased further, there was an abrupt decrease in $T_c$ which shifts down to 140 K at 171 GPa.

The pressure dependence of $T_c$ in our two experiments is summarized in Fig. 3, where we have plotted our new data together with all previous experimental measurements and the results from our previous theoretical calculations (*6-9*). The errors in the measured critical temperature and pressure are also displayed. The error in the pressure is estimated from the difference between the initial and final pressures during each temperature cycle, while the error in the temperature is estimated from the difference between the onset and completion temperatures. The uncertainties in determining $T_c$ are especially large above 150 GPa. To reach ultrahigh pressure the diameter of the diamond anvil culets is reduced, leading to a reduction of the magnetic signal which is proportional to the volume of the sample. The signal-to-noise ratio is thus reduced, resulting in the larger uncertainty in $T_c$ upon further compression.

From the plot of $T_c$ versus pressure in Fig. 3 it is clearly seen that the diagram can be divided into two regions. Below 140 GPa the onset temperature $T_c$ is lower than 100 K, which was also observed in previous experiments reported by Eremets *et al.* (*7*). Near 149 GPa high $T_c$ superconductivity appears and simultaneously $H_3S$ becomes the dominant component in the sample, as shown by the reported X-ray diffraction data (*9*). It should be noted that there is a difference between the critical temperatures deduced from electrical resistance data and from our magnetic susceptibility measurement, and this difference increases by 30 K with increasing pressure up to 170 GPa. Still, there is no doubt that we have identified high temperature superconductivity in the sulfur hydride system from its Meissner effect. A quick overview of previous superconductivity experiments under high pressure also shows that the superconducting critical temperature observed in electrical resistance measurements is always higher than that found from magnetic susceptibility data (*13*, *14*). Part of this difference could be due to the fact that the magnetic measurements are sensitive to the bulk fraction of superconductive material only, while the resistance measurements may also react to small amounts of superconducting material inside or between grains, creating percolation paths through the material.

A close study of Fig. 3 shows that there is a pronounced kink in the dependence of $T_c$ on pressure at about 149 GPa. This was also observed in the reported electrical resistance measurements (*7*, *9*). Our earlier theoretical calculation could give the rational explanation for this kink (*6*), since the predicted *R*3*m* and *Im*-3*m* phases have different pressure dependences of $T_c$. However, the *R*3*m* and *Im*-3*m* phases cannot be distinguished from the powder X-ray diffraction patterns (*9*), since these two crystal structures differ only by a single different hydrogen atom position. Further neutron scattering studies are necessary to confirm the transition between the *R*3*m* and *Im*-3*m* phases.

In summary, the excellent superconducting properties of compressed sulfur hydride have been confirmed by magnetic susceptibility measurements. With increasing pressure, superconductivity appears at 38 K for 117 GPa, $T_c$ rises rapidly to 183 K at 149 GPa and then decreases to 140 K at 171 GPa. The pressure dependence of $T_c$ suggests a phase transition, which is consistent with both theoretical calculations and electrical resistance measurements. The present results, showing a true Meissner effect, provide enormously important evidence for a superconducting state with extremely high critical temperature in hydrogen sulfide. This should stimulate a great deal of future experimental and theoretical studies to explore possible high-temperature superconductivity in other hydrogen-dominated materials.

**References**

1. H. Kamerlingh Onnes, Further experiments with liquid helium. C. On the change of electric resistance of pure metals at very low temperatures, etc. IV. The resistance of pure mercury at helium temperatures. *Commun. Phys. Lab. Univ. Leiden* No. 120b (1911).
2. N. W. Ashcroft, Metallic Hydrogen: A High-Temperature Superconductor? *Phys. Rev. Lett.* **21,** 1748 (1968).
3. N. W. Ashcroft, Hydrogen Dominant Metallic Alloys: High Temperature Superconductors? *Phys. Rev. Lett.* **92**, 187002 (2004).
4. X. Jin, X. Meng, Z. He, Y. Ma, B. Liu, T. Cui, G. Zou, H. - K. Mao, Superconducting high-pressure phases of disilane. *Proc. Natl. Acad. Sci U S A* **107**, 9969-9973 (2010).
5. H. Wang, J. S. Tse, K. Tanaka, T. Iitaka, Y. Ma, Superconductive sodalite-like clathrate calcium hydride at high pressures. *Proc. Natl. Acad. Sci U S A* **109**, 6463-6466 (2012).
6. D. Duan, Y. Liu, F. Tian, D. Li, X. Huang, Z. Zhao, H. Yu, B. Liu, W. Tian, and T. Cui, Pressure-induced metallization of dense $(H_2S)_2H_2$ with high-$T_c$ superconductivity. *Sci. Rep.* **4,** 6968 (2014).
7. A. P. Drozdov, M. I. Eremets, and I. A. Troyan, Conventional superconductivity at 190 K at high pressures. Preprint at http://arxiv.org/abs/1412.0460 (2014).
8. A. P. Drozdov, M. I. Eremets, I. A. Troyan, V. Ksenofontov, and S. I. Shylin, Conventional superconductivity at 203 kelvin at high pressures in the sulfur hydride system. *Nature* **525,** 73–76 (2015).
9. M. Einaga, M. Sakata, T. shikawa, K. Shimizu, M. I. Eremets, and A. P. Drozdov, I. A. Troyan, N. Hirao and Y. Ohishi, Crystal structure of the superconducting phase of sulfur hydride. *Nature Physics* **12**, 835-838 (2016).
10. D. Duan, X. Huang, F. Tian, D. Li, H. Yu, Y. Liu, Y. Ma, B. Liu and T. Cui, Pressure-induced decomposition of solid hydrogen sulfide. *Phys. Rev. B* **91**, 180502 (2015).
11. I. Troyan, A. Gavriliuk, R. Rüffer, A. Chumakov, A. Mironovich, I. Lyubutin, D. Perekalin, A. P. Drozdov, and M. I. Eremets, Observation of superconductivity in hydrogen sulfide from nuclear resonant scattering. *Science* **351**, 1303-1306 (2016).
12. Y. A. Timofeev, V. V. Struzhkin, R. J. Hemley, H. - K. Mao, and E. A. Gregoryanz, Improved techniques for measurement of superconductivity in diamond anvil cells by magnetic susceptibility. *Rev. Sci. Instrum.* **73,** 371 (2002).
13. S. Deemyad, and J. S. Schilling, Superconducting Phase Diagram of Li Metal in Nearly Hydrostatic Pressures up to 67 GPa. *Phys. Rev. Lett.* **91**, 167001 (2003).
14. K. Shimizu, H. Ishikawa, D. Takao, T. Yagi, and K. Amaya, Superconductivity in compressed lithium at 20 K. *Nature* **419**, 597-599 (2002).

**Acknowledgments**

This work was supported by National Natural Science Foundation of China (Nos. 11504127, 51572108, 51632002, 11634004, 11274137, 11474127), Program for Changjiang Scholars and Innovative Research Team in University (No. IRT_15R23), National Found for Fostering Talents of basic Science (No. J1103202), and China Postdoctoral Science Foundation (2015M570265).

**Supporting Online Material**

Methods
Supplementary Text
Fig. S1
Reference (15)

**FIGURES**

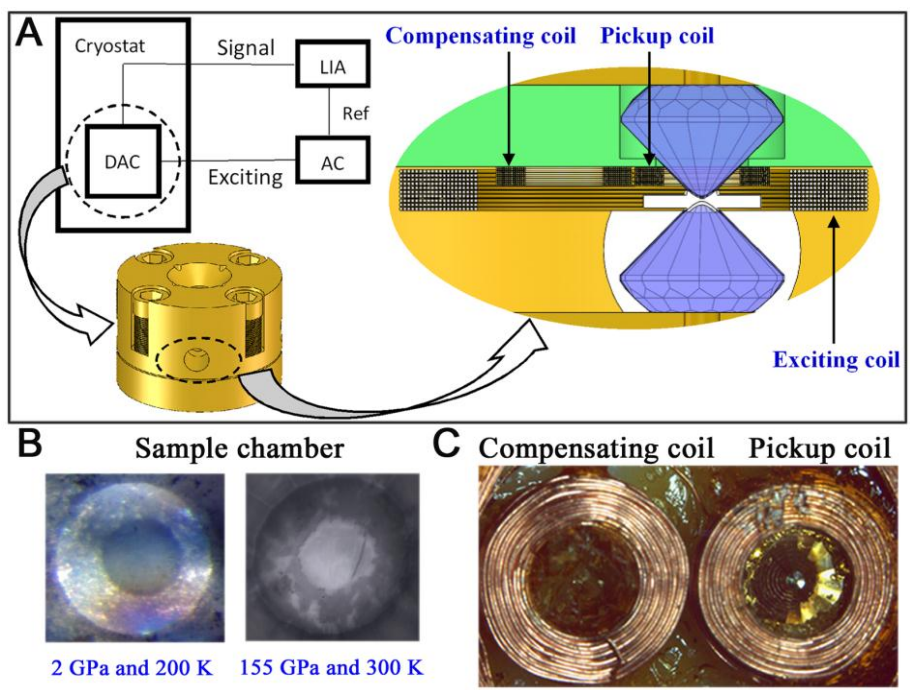

**Fig. 1.** Schematic image of the experimental setup for alternating-current magnetic susceptibility measurement. A. Simplified flow chart for the magnetic susceptibility measurement set-up for the diamond anvil cell. LIA and AC denote lock-in amplifier and alternating-current source, respectively. B. The sample in the gasket hole at 2 GPa and 200 K (left), and at 155 GPa and 300 K (right), respectively. C. A pickup coil wound around a diamond anvil and a compensating coil connected in opposition.

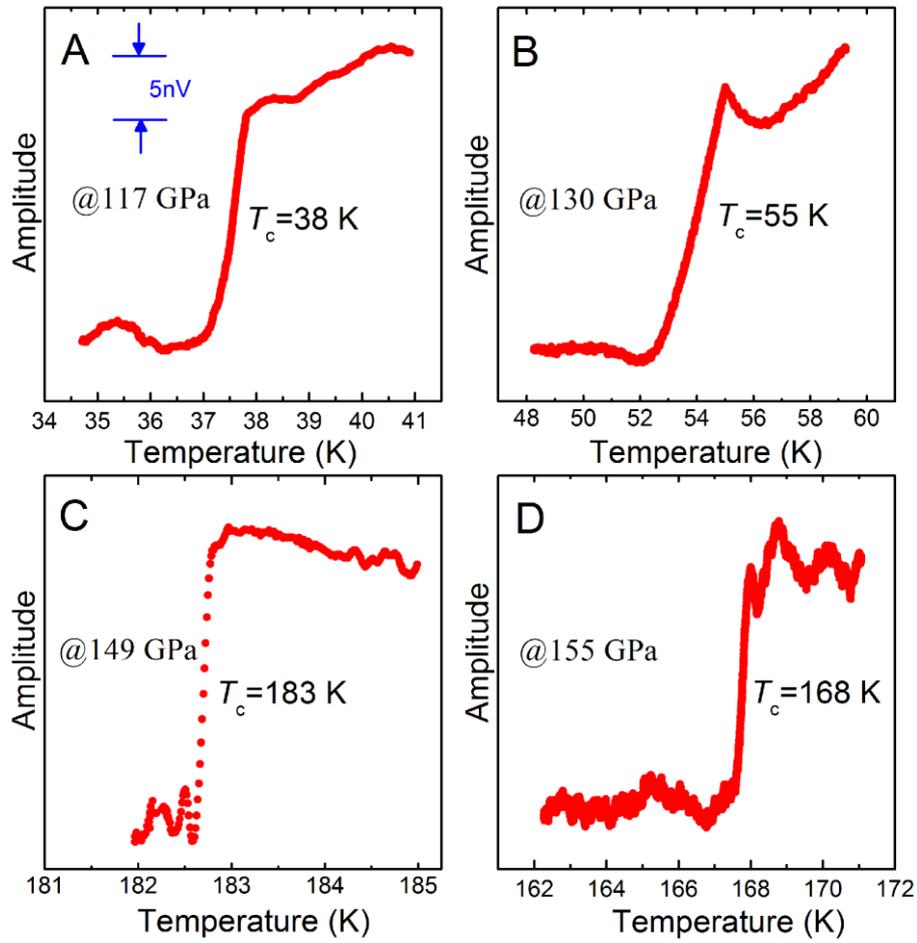

**Fig. 2.** Magnetic susceptibility signals of sulfur hydride at various pressures. Curves show typical amplitude records from the coil system during each temperature scan. To obtain the strong amplitude signal, we kept the sample and compensating coil signals in phase. The critical temperature $T_C$ is estimated from the onset of the superconducting transition.

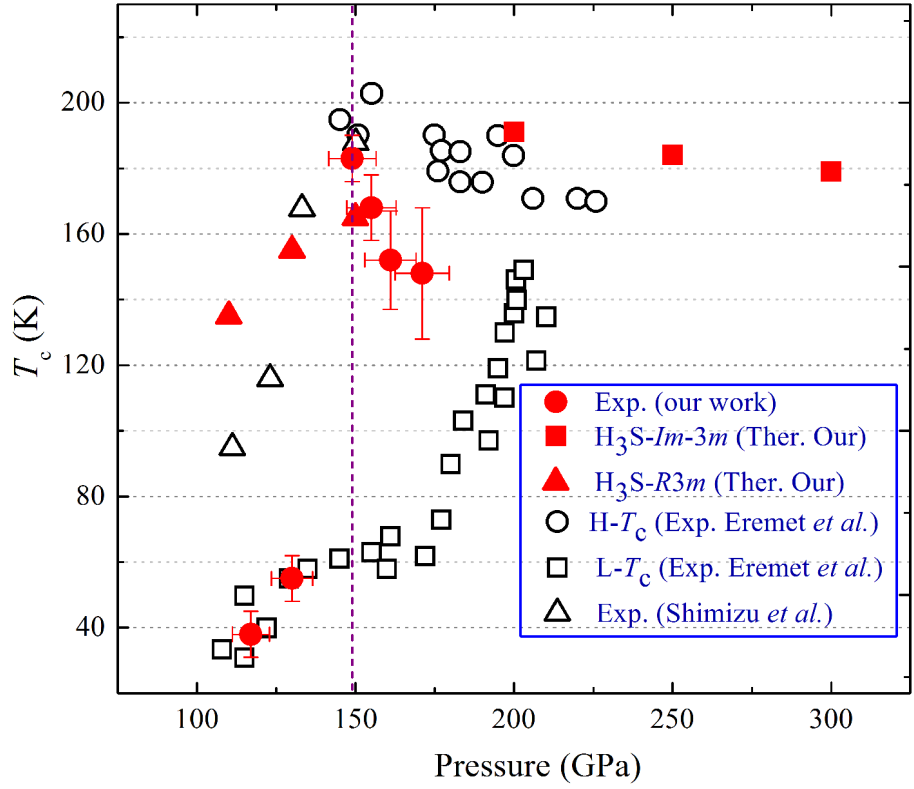

**Fig. 3.** Pressure dependence of $T_c$ in sulfur hydride. Filled circles show data from this experiment while filled squares and triangles indicate our reported theoretical work (*6*). The open circles, squares and triangles show experimental results reported by Eremet *et al.* (*7, 8*) and Shimizu *et al.* (*9*), respectively.

# Supporting Online Material

## METHODS

**Sample preparation:** For ultrahigh pressure generation, diamond anvil cells (DACs) with anvil culets of 100–200 $\mu m$ have been used. Commercial gaseous $H_2S$ with purity 99.999% was loaded into DACs with rhenium gaskets in a argon glove box. For the loading of $H_2S$ gas into the DACs, all tubes from the lecture bottle to the DAC were first purged of a small amount of the gas, and a fresh $H_2S$ sample was used to minimize the contaminations. The $H_2S$ sample was usually condensed by spraying it as vapor into the gasket hole of a DAC cooled in liquid nitrogen. When the hole is full of solidified $H_2S$, the upper diamond is translated to seal the sample by increasing pressure. After clamping the sample it was compressed to at least 100 GPa at temperatures $T < 100$ K to avoid possible decomposition. Pressure at low and room temperatures was determined from the sharp edge of the first-order diamond Raman band produced at the diamond–sample surface of the stressed anvil (*1*). Temperature was measured with a calibrated Si-diode attached to the DAC.

The Raman spectra were collected with a triple JYT64000 micro-Raman system. A solid-state, diode laser pumped, frequency-doubled Nd: vanadate laser with a wavelength of 532 nm was used as excitation laser.

**Magnetic susceptibility.** For magnetic measurements, we used diamond anvil cells fabricated from Cu-Be alloy. The susceptibility of the metallic parts of the high-pressure cell is essentially independent of the external field. Applying a magnetic field to the cell when it contains a sample will therefore change the signal coming from the sample whereas the background arising from the other surrounding parts will remain nearly constant. This allows the signal from the sample to be separated from that coming from the background. The signal changes abruptly in the vicinity of the superconducting transition, allowing us to see these changes in the amplitude of the signal.

**Reference**

15. M. I. Eremets, Megabar high-pressure cells for Raman measurements. *J. Raman Spectrosc.* **34**, 515–518 (2003).

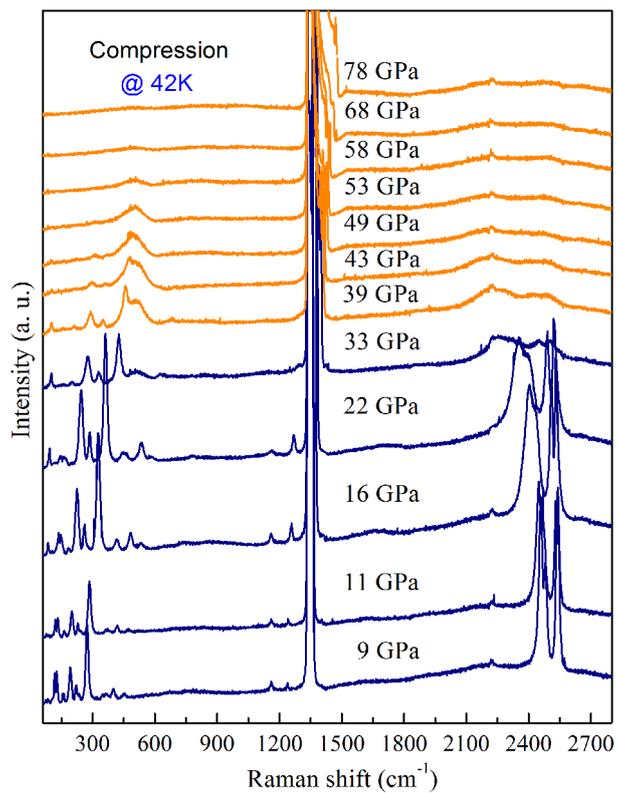

**Fig. S1.** Raman spectra of sulfur hydride measured at different pressures at about 42 K. At 39 GPa, there is a phase transition with the disappearance of characteristic peaks.